\documentclass[letterpaper]{sfchem}
\usepackage{graphicx}
\def\hhcs{H$_2$CS}
\def\hhs{H$_2$S}

\def\soo{SO$_2$}

\def\gtsim{{_>\atop{^\sim}}}
\def\ltsim{{_<\atop{^\sim}}}

\def\lsol{L$_{\odot}$}

\def\txc{$T_{\rm ex}$}

\begin{document}

\title{Sulphur chemistry in the envelopes of massive young stars}

\author{F.F.S. van der Tak \inst{1} \and A.M.S. Boonman \inst{2}\and
  R. Braakman \inst{2}\and E.F. van Dishoeck\inst{2}} 

\institute{Max-Planck-Institut f\"ur Radioastronomie, Auf dem H\"ugel
  69, 53121 Bonn, Germany \and Sterrewacht, Postbus 9513, 2300 RA
  Leiden, The Netherlands} 

\authorrunning{van der Tak et al.}
\titlerunning{Sulphur chemistry around massive young stars}

\maketitle 

\begin{abstract}
  
  We present submillimeter observations of SO, \soo, \hhs, \hhcs, OCS,
  NS and HCS$^+$ toward nine massive young stars.  The outflow
  contributes $\approx$50\% to the SO and \soo\ emission in $15-20''$
  beams, more than for CS, where it is 10\%.  The \soo\ abundance
  increases from dark cloud levels in the outer envelope ($T<100$~K)
  to levels seen in hot cores and shocks in the inner envelope
  ($T>100$~K).  Molecular abundances are consistent with a model of
  ice evaporation in an envelope with gradients in temperature and
  density for a chemical age of $\sim 30000$~yr. The high observed
  abundance of OCS, the fact that \txc (OCS)$\gg$\txc (\hhs), and the
  data on solid OCS and \hhs\ all suggest that the major sulphur
  carrier in grain mantles is OCS rather than \hhs.  For most other
  sulphur-bearing molecules, the source-to-source abundance variations
  by factors of up to 10 do not correlate with previously established
  evolutionary trends in temperature tracers.  These species probe the
  chemically inactive outer envelope.  Our data set does not constrain
  the abundances of \hhs\ and SO in the inner envelope, which,
  together with \soo, are required to use sulphur as a clock.


\end{abstract}

\section{Introduction}
\label{sec:intro}

Spectral line surveys at submillimeter wavelengths have revealed
considerable chemical differences between star-forming regions (see
\cite{evd01} for an 
overview). While these differences indicate activity, the
dependence of molecular abundances on evolutionary state and physical
parameters is poorly understood. Better insight into this relation
would be valuable for probing the earliest, deeply embedded phases of
star formation where diagnostics at optical and near-infrared
wavelengths are unavailable. This is especially true for the formation
of high-mass stars, for which the order in which phenomena occur is
much less well understood than in the low-mass case, and for which the
embedded phase is a significant fraction ($\approx$10\%) of the total
lifetime of $\sim$10$^6$~yr.

\begin{figure}
  \begin{center}
\resizebox{\hsize}{!}{\includegraphics[height=10cm,angle=-90]{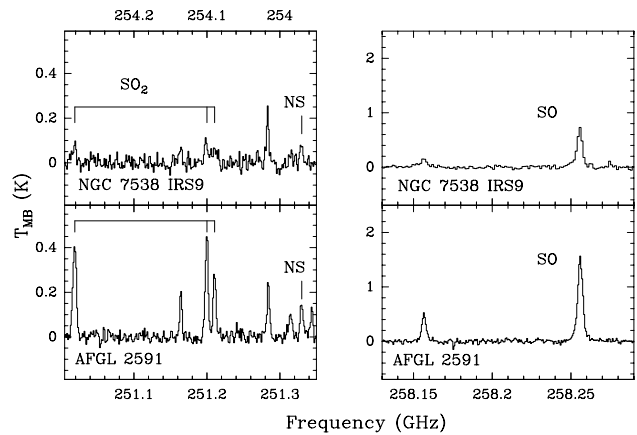}}      
    \caption{Examples of the spectra 
      taken with the JCMT. The top and bottom frequency scales of the
      left panel are the two receiver sidebands; the right panel only
      shows lines from one sideband. Adopted from van der Tak et al.~(2002).}
    \label{fig:data}
  \end{center}
\end{figure}

Sulphur-bearing molecules are attractive as candidate tracers of early
protostellar evolution (\cite{jh98sulf}). Models of ``hot
core'' chemistry (\cite{char97}), as well as of shock chemistry
(\cite{gpdf93}) predict strong variations in the abundances of sulphur
molecules on time scales of $\sim$10$^4$~yr. To test these scenarios,
this paper discusses submillimeter observations of sulphur-bearing
molecules towards nine regions at the earliest stages of high-mass
star formation, with luminosities $1\times 10^4 - 2\times 10^5$~\lsol\
at distances of $1-4$~kpc. This paper only summarizes our results; a
complete description is given by \cite*{fvdt02}.

\section{Observations}
\label{sec:obs}

\begin{figure}
  \begin{center}
\resizebox{\hsize}{!}{\includegraphics[angle=0]{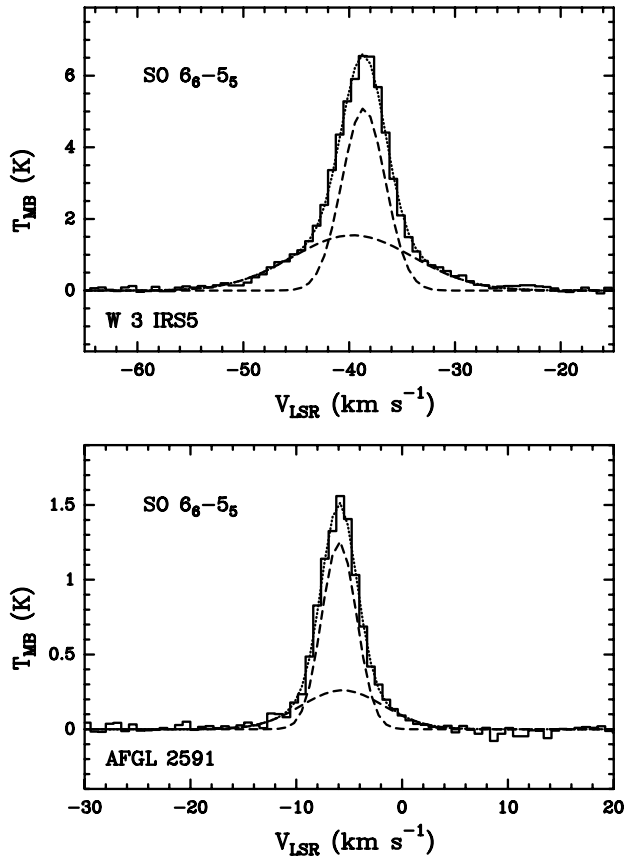}}    
    \caption{Velocity profiles of the SO 6$_6$--5$_5$
      line in the sources AFGL~2591 and W3~IRS5. }
    \label{fig:2comp}
  \end{center}
\end{figure}

Spectra in the 230 and 345~GHz bands were taken in 1995--1998 with the
JCMT.  The overview spectra in Fig.~\ref{fig:data} show
source-to-source differences in the relative and in the absolute
strengths of lines of sulphur molecules. The line profiles
(Fig.~\ref{fig:2comp}) reveal the presence of a narrow component, due
to the envelope, and a broad one, due to the outflow. For SO and \soo,
the outflow contributes $\approx$50\% of the line flux. This fraction
is much larger than for CS, where it is only 10\%. Lines of other
molecules are too weak to separate the two components.

\begin{figure}
  \begin{center}
\resizebox{\hsize}{!}{\includegraphics[angle=0]{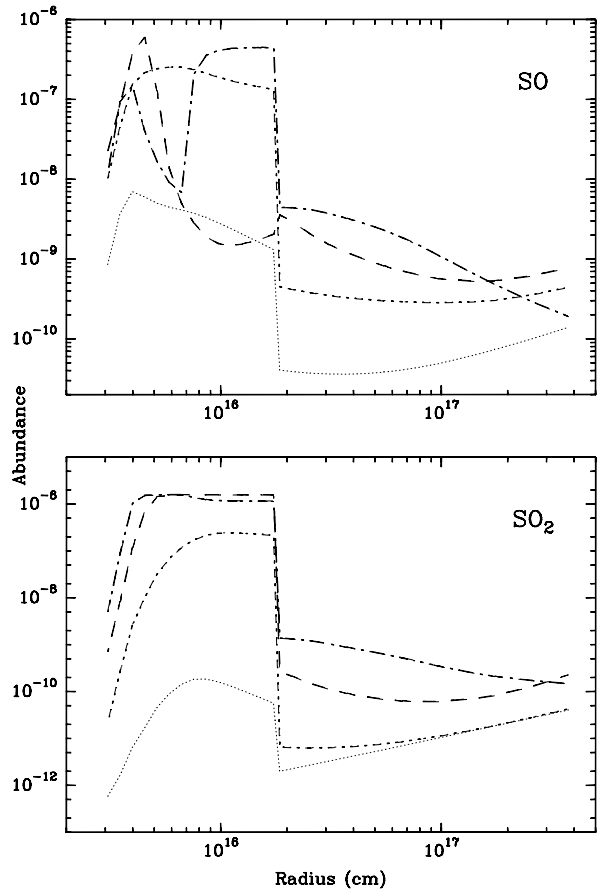}}        
    \caption{Distributions of the SO and \soo\
      abundances in AFGL 2591 for chemical ages of 300, 3000, 30000
      and 300000 yr (curves from bottom to top) after Doty et al.~(2002).
      The $t\sim 3\times 10^4$~yr model fits our data best.}
    \label{fig:doty}
  \end{center}
\end{figure}

\section{Results}
\label{sec:hst}

To calculate molecular abundances, we have used the Monte Carlo
program by \cite*{hst00}, and models of the temperature and density
structure of the sources (\cite{fvdt00}). Table~\ref{t:abs} compares
our results with abundances in other types of sources.  The best
overall match is with ``hot cores''.

For some sources, the detection of many SO$_2$ lines from a wide range
of energy levels enables us to derive radial abundance profiles.
Models with a constant SO$_2$ abundance of $10^{-9}$ fit the lines
arising $\ltsim 100$~K above ground.  However, the higher-excitation
lines are underproduced, which indicates a `jump' in the SO$_2$
abundance by a factor of $\sim$100 at a temperature of $\approx
100$~K.  Infrared observations of SO$_2$ absorption by \cite*{kean01}
also indicate abundances of $\sim$10$^{-7}$ for the inner envelopes.

\begin{table*}
\caption{Molecular abundances.}
\label{t:abs}
  \begin{center}
    \begin{tabular}[lrrrrrrrr]{lrrrrrrrr}
\hline \hline
\noalign{\smallskip}
Source  & CS      & H$_2$CS  & H$_2$S  & HCS$^+$  & NS       & OCS     & SO      & SO$_2$ \\
        &$10^{-9}$&$10^{-10}$&$10^{-9}$&$10^{-10}$&$10^{-11}$&$10^{-9}$&$10^{-9}$&$10^{-9}$\\
\noalign{\smallskip}
\hline
\noalign{\smallskip}
W3 IRS5       &  5 &  3  &  9    & 0.2 & ...   & 0.5 & 10 & 10  \\
W33A          &  5 & 13  &  8    & 2   & 2     & 20  & 1  & 1   \\
AFGL 2136     &  4 &  3  & $<$2  & 1   &$<$0.5 & ... & 1  & 0.5 \\
AFGL 2591     & 10 &  3  &  20   & 2   & 0.5   & 10  & 10 & 2   \\
S140 IRS1     &  5 &  7  &  4    & 5   & 1     &$<$2 & 2  & 1   \\
NGC 7538 IRS1 & 10 &  10 &  8    & 2   & 1     & 2   & 5  & 1   \\
NGC 7538 IRS9 & 10 &  10 &  8    & 5   & 1     & 2   & 1  & 0.5 \\
MonR2 IRS3    &  5 &  7  & $<$4  & ... & ...   & ... & 5  & 1   \\
NGC 6334 IRS1 & 10 &  7  &  10   & 2   & 10    & 50  & 2  & 2   \\
\noalign{\smallskip}
\hline
\noalign{\smallskip}
Hot cores$^a$  &  8 &  6  & 9     & ... & 80    &  5  & 4 & 20  \\
PDRs$^b$       & 20 & ... & 6     & ... & ...   & ... & 9 & 0.1 \\
Dark clouds$^c$&  1 &  6  & 0.8   & 0.6 & ...   &  2  & 20& 4   \\
Shocks$^d$     &  4 &  8  & 4000  & 0.1 & ...   & 10  &200& 100 \\
\noalign{\smallskip}
\hline
    \end{tabular}

$^a$ Source average from \cite*{jh98sulf}
$^b$ Orion Bar: \cite*{jansen95}

$^c$ L~134N: \cite*{ohishi92}
$^d$ Orion Plateau: \cite*{sutton95,minh90}

%$^e$ Increased by a factor of $\sim$300 in the inner envelope ($T>100$~K).

  \end{center}
\end{table*}

\section{Chemical implications}
\label{sec:doty}

Most hot core models only consider single values for the temperature
and density, while for our sources, these parameters vary strongly
along the line of sight. Using the UMIST database, \cite*{doty02}
calculated the time-dependent gas-phase chemistry for the temperature
and density structure of AFGL 2591. The model has most sulphur initially
in S at $T<100$~K and in H$_2$S at $T>100$~K, to mimic the effect of
ice evaporation. The resulting `jump' in the abundances at $r=2\times
10^{16}$~cm is  consistent with our observations of \soo. See
Fig.~\ref{fig:doty} and Doty (this volume).

The model reproduces most of our observed abundances, and also the
increase of the \soo\ abundance at $T\gtsim100$~K. However, OCS is
underproduced by factors of 10--100, suggesting that most of the
sulphur in the grain mantles is in the form of OCS. This idea is
supported by the high excitation temperature of OCS (100~K), much
higher than the 25~K found for H$_2$S. In addition, solid OCS has been
detected in the infrared towards W33A, while H$_2$S has not.

The current data set is not suitable to trace chemical evolution
within our sample. The CS, \hhcs, HCS$^+$ and NS molecules mainly
probe the chemically inactive outer envelope ($T<100$~K).  While OCS,
SO and \soo\ show larger source-to-source variations, only for \soo,
the data allow to determine the abundance in the inner envelope
($T<100$~K) where ices are evaporating.  The situation for \hhs\ is
unclear: models predict strong abundance variations, but the present
data do not trace the warm gas.  Future observations of
high-excitation lines of \hhs, SO and OCS could exploit the full
potential of sulphur as chemical clock.

\bibliographystyle{aa}
\bibliography{vdtak_poster}
\end{document}